\documentclass[letter]{aa} 
\usepackage{graphicx}
\usepackage{txfonts}
\usepackage{color}

\begin{document}
   \title{Optical identification of the transient supersoft X-ray source RX J0527.8-6954, in the LMC
	\thanks{Based on observations obtained at the Gemini Observatory, which is operated by the
Association of Universities for Research in Astronomy, Inc., under a cooperative agreement
with the NSF on behalf of the Gemini partnership: the National Science Foundation (United
States), the Science and Technology Facilities Council (United Kingdom), the
National Research Council (Canada), CONICYT (Chile), the Australian Research Council
(Australia), Minist\'erio da Ci\^encia e Tecnologia (Brazil) and SECYT (Argentina).}
	}

   \author{A. S. Oliveira \inst{1}
  \and
  J. E. Steiner \inst{2} \and T.V. Ricci \inst{2} \and R.B. Menezes \inst{2} \and B.W. Borges \inst{2}
          }

   \institute{IP\&D, Universidade do Vale do Para\'{\i}ba, Av. Shishima Hifumi, 2911, CEP 12244-000, S\~ao Jos\'e dos 
Campos, SP, Brasil \\
	     \email{alexandre@univap.br}
	     \and
	     Instituto de Astronomia, Geof\'{\i}sica e Ci\^encias Atmosf\'ericas, Universidade de S\~ao Paulo, 05508-900,
    S\~ao Paulo, SP, Brasil\\
             }

   \date{Received / Accepted}

 
  \abstract
   {Close binary supersoft X-ray sources (CBSS) are binary systems that contain a white dwarf  with stable nuclear burning on its surface. These sources, first discovered in the Magellanic Clouds, have high accretion rates and near-Eddington luminosities ($10^{37}-10^{38}$ erg~s$^{-1}$) with high temperatures ($T=2-7 \times 10^{5}$ K).}
   {The total number of known objects in the MC is still small and, in our galaxy, even smaller.  We observed the field of the unidentified transient supersoft X-ray source \object{RX J0527.8-6954} in order to identify its optical counterpart.}
   {The observation was made with the IFU-GMOS on the Gemini South telescope with the purpose of identifying stars with possible \ion{He}{ii} or Balmer emission or else of observing  nebular extended jets or ionization cones, features that may be expected in CBSS.}
   {The X-ray source is identified with a B5e V star that is associated with subarcsecond extended H$\alpha$ emission, possibly bipolar.}
   {If the primary star is a white dwarf, as suggested by the supersoft X-ray spectrum, the expected orbital period exceeds 21 h; therefore, we believe that the 9.4 h period found so far is not associated to this system.}

   \authorrunning{A. S. Oliveira et al.}
   \titlerunning{Optical identification of RX J0527.8-6954}
   \maketitle
%

\section{Introduction}

Close binary supersoft X-ray sources (CBSS) are binary systems in which a white dwarf displays hydrostatic nuclear burning on its surface (van den Heuvel et al. \cite{heuvel}). This happens because the accretion rate is very high ($\dot{M}\sim10^{-7}$ M$_{\odot}~$yr$^{-1}$). As a consequence, the luminosity of these systems reaches high values ($10^{37}-10^{38}$ erg s$^{-1}$), and the temperature of the radiation is $2-7 \times 10^5$ K. Most of the radiation is emitted in the ultraviolet and supersoft X-ray bands. The first sources were discovered in the Magellanic Clouds (MC) and identified as a class by the ROSAT satellite (Tr\"umper et al. \cite{trumper}). Up to now only two sources have been detected in  the Galaxy (see Kahabka \& van den Heuvel \cite{kah}, for a review and references).

The question of why there is such a discrepancy between the MC and the Galaxy has been addressed by Steiner \& Diaz (\cite{stei98}), who proposed that the V Sge stars are the galactic counterparts of the CBSS not detected in X-rays. This could happen if the supersoft emission is absorbed by the interstellar gas, much denser and with much higher metallicity in the Galaxy than in the MC.

If the source emits copious radiation that is not detected directly, it could still be noticed if the local environment is photoionized by the radiation. There are enough ionizing photons in these systems to do this, but is there sufficient gas for this to be noticed? All CBSS and V Sge stars seem to present strong winds, escaping from the white dwarf, disk, and even the secondary in many cases (van Teeseling \& King \cite{tees}). Other CBSS and V Sge stars produce jets, seen as spectral satellites to the Balmer and \ion{He}{ii} lines (see Steiner et al. \cite{stei07} for references). A significant amount of gas must exist in the immediate environment, perhaps collimated in the form of jets, perpendicular to the disk. In systems seen pole-on (low inclination), this could be noticed as a nebulosity. In systems with high inclination, this would be noticed as extended nebular emission with bipolar geometry. This could have shape of jets or, perhaps, ionization cones (seen also in AGN). Attempts to detect nebular emission from CBSS/V Sge have failed, with the significant exception of \object{CAL~83} (Remillard et al. \cite{remi}). But in that case, the star is located near a massive cloud that has little, if any, to do with the star.

\object{RX J0527.8-6954} was discovered as an LMC supersoft X-ray source by ROSAT observations (Tr\"umper et al. \cite{trumper}) with spectral parameters similar to those of the CBSS prototype CAL~83 (Greiner et al. \cite{greiner91}). It has not been detected by previous Einstein observations of the field, suggesting that it was at least $\sim 10$ times fainter by that time. In addition, its X-ray flux decreased by a factor of $\sim 50$ from its discovery in 1990 to ROSAT observations taken in 1995 (Greiner et al. \cite{greiner96a}).  RX J0527.8-6954 is, thus, a special case among the CBSS that concern the X-ray variability.

\begin{figure}
      \resizebox{\hsize}{!}{\includegraphics[clip]{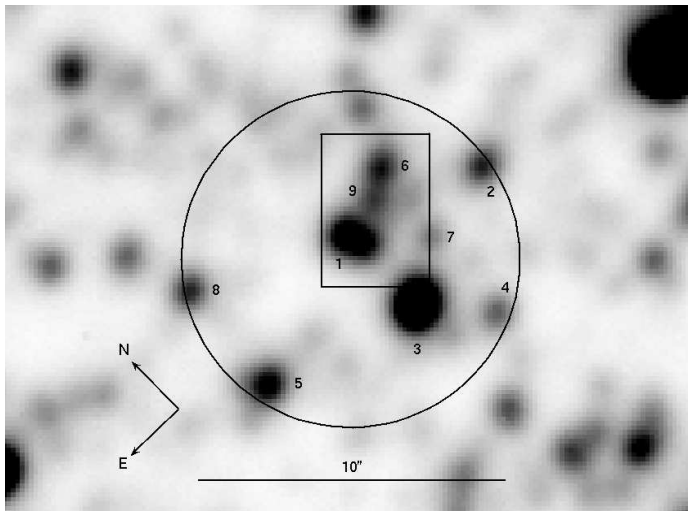}}
           \caption{GMOS-IFU acquisition image, with stars 1--9 as defined by Greiner \& Hazen (\cite{greiner96}). The ROSAT 
$5\arcsec$ error circle and the IFU $5\arcsec \times 3.5\arcsec$ field of view are also shown.}
     \label{fig1}
  \end{figure}

The optical counterpart of this object, on the other hand, has never been confidently identified. The ROSAT $5\arcsec$  error circle encompasses at least 9 stars that could be associated to the CBSS. Greiner et al. (\cite{greiner96b}) and Greiner \& Hazen (\cite{greiner96}) numbered all resolved objects inside this error circle, and we adopt those identification numbers in this work (see Fig.~\ref{fig1}). The history of the counterpart identification has often approached the elusive Harvard variable star \object{HV 2554}, but this connection seems to have been lost, since the optical variability of HV 2554 was set as doubtful (Greiner \& Hazen \cite{greiner96}) and the improved X-ray position of  RX J0527.8-6954 put HV 2554 outside the X-ray error circle (Cowley et al. \cite{cow97}). Greiner \& Hazen (\cite{greiner96}) stated the lack of optical variability of stars 1 to 9 (with some doubt for the fainter objects), suggesting that none of them are the optical counterpart of  RX J0527.8-6954. Cowley et al. (\cite{cow97}) determined spectral type B8 IV for star 1 and possibly G giant for star 3, while star 2 is supposedly a B type star (Greiner et al. \cite{greiner96b}). Using MACHO project photometry, Alcock et al. (\cite{alco}) attempted to optically identify the source. From the analysis of the lightcurves of all objects 1--9, they proposed a steady decline of 0.51 mag in the blue for stars 6 or 9 over 4.25 years. After removing this long term decline, a sinusoidal variation  with period of 9.42 h and 0.052 mag amplitude remained. They suggest, therefore, that this optical variability is associated to the CBSS.
In this paper, we show that star 1 is a blend of two stars and report the optical identification of RX J0527.8-6954 by detecting extended nebular emission in H$\alpha$, with a bipolar subarcsecond structure associated to the brightest of the pair.

\section{Observations and data reduction}

The data on RX J0527.8-6954 were obtained in queue mode on September 2004 on the Gemini South Telescope with the Gemini Multi Object Spectrograph (GMOS -- Hook et al. \cite{hook}, Allington-Smith et al. \cite{alli}), operated in the integral field unit (IFU) mode. The GMOS-IFU has two fields of view that simultaneously sample both the target and a contiguous region of sky, displaced by 1 arcmin from the object field, by closed packed arrays of 1000 and 500 lenslets, respectively. Each lenslet feeds light to a coupled optical fiber, which sends the light to a linear array at the nominal location of the slit of the spectrograph, the pseudo slit. The final data are three-dimensional (two spatial and one spectral) data cubes that allow either the construction of narrow-band images of any desired bandwidth at any slice of the spectral dimension or extraction of a spectrum from any point in the field of view. We set the GMOS-IFU to operate with one pseudo-slit, which resulted in greater spectral coverage at the expense of field of view  area (sampled by 500 lenslets in this configuration). The adopted B600\_G5323 grating yielded average spectral resolution R=2900 in the range from 4100 {\AA} to 6900 {\AA} over an FOV of $5\arcsec \times 3.5\arcsec$. Three 15-minute exposures were obtained and summed after reduction. Lamp calibration flats, twilight flats, CuAr arc exposures, and bias images were also taken to calibrate the data.
The seeing of the observation was $0.6\arcsec$.

The data reduction was performed with the IRAF\footnote{IRAF is distributed by the National Optical Astronomy Observatories, which are operated by the Association of Universities for Research in Astronomy, Inc., under cooperative agreement with the National Science Foundation.} $gemini.gmos$ package, comprising bias and background subtraction, cosmic ray rejection, spectra extraction, wavelength calibration, CCD pixel sensitivity and fiber response correction, sky subtraction, and data cube construction. The final data cubes have $0.05\arcsec$ spatial sampling. 
As the Atmospheric Dispersion Corrector (ADC) at GMOS is not functional, the differential atmospheric refraction implies that the data have a spatial distortion over the spectral range, especially at high airmass. To deal with this problem, we implemented an algorithmic procedure that corrects this distortion.

\section{Data analysis and results} 

\begin{figure}
      \resizebox{\hsize}{!}{\includegraphics[clip]{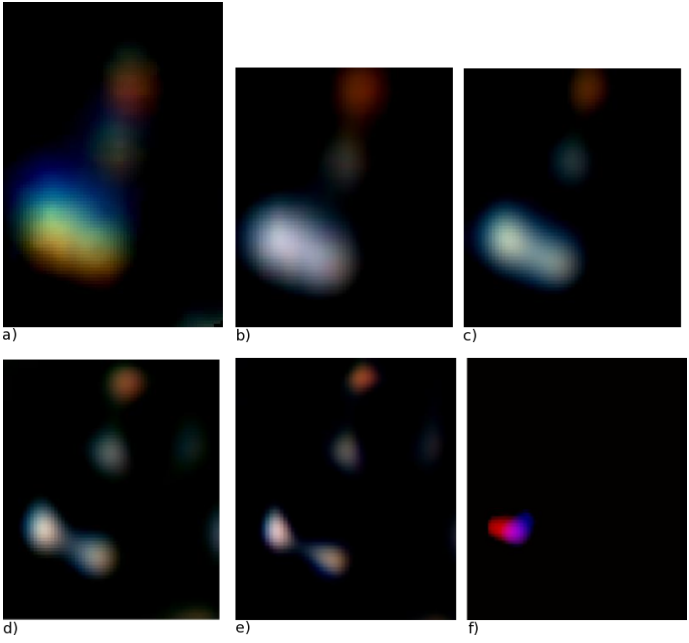}}
           \caption {\textit{a)} The original data cube divided in three equal segments and displayed in RGB true colors, 
showing stars 1, 9, and 6. The rainbow aspect is introduced by differential atmospheric refraction. \textit{b)} The same as a) after Butterworth-filtering in Fourier space and correcting for atmospheric differential refraction. \textit{c)} The same as b) after Lucy- Richardson deconvolution with a Gaussian PSF with FWHM$=0.6\arcsec$. \textit{d)} The second wavelet component of the cube is shown in c). \textit{e)} The sum of the first and second wavelet components, after Landweber deconvolution. Star 1 is clearly separated in two components: star 1a (the brightest at the left) and star 1b. Star 9 has a PSF with FWHM$=0.37\arcsec$. \textit{f)} Blue and red wings of H$\alpha$ emission after subtracting the adjacent continuum.}
     \label{fig2}
  \end{figure}

After the data cube (Fig.~2a) was corrected for differential atmospheric refraction, we followed noise reduction and deconvolution procedures that are fully described and discussed in Steiner et al. (in preparation). Basically the following steps were adopted and illustrated in Fig.~\ref{fig2}.

\begin{itemize}
 \item We transformed the data cube to Fourier space so that the spatial dimensions ($x,y$) are represented in spatial frequencies ($u,v$).

\item A Butterworth filter was applied to remove high spatial frequency noise.

\item The inverse Fourier transform was made (Fig.~2b) and Richardson-Lucy deconvolution was applied (Fig.~2c), using a Gaussian PSF with FWHM$=0.6\arcsec$ and 6 iterations. 

\item A wavelet transform was applied to the spatial dimension of the data cube, creating 6 ``wavelet cubes'' (Fig.~2d shows the wavelet of second highest frequency component).

\item A 30-iteration Landweber deconvolution was applied to each wavelet cube. Figure ~2e shows the first two wavelet components.

\item The blue and red wings of H$\alpha$ emission were imaged, after subtracting the adjacent continuum (Fig.~2f). 

\end{itemize}

\begin{figure}
      \centering {\includegraphics[width=0.35\textwidth]{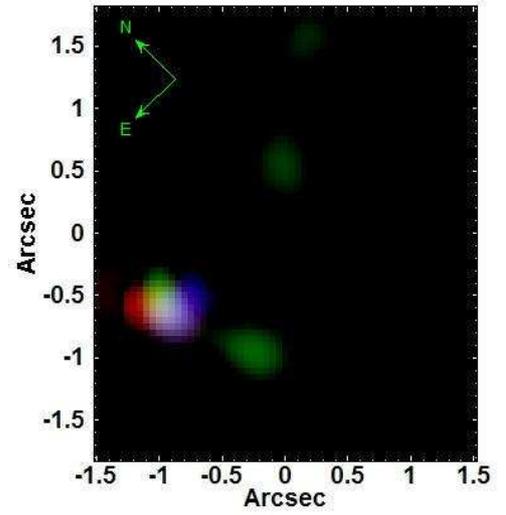}}
           \caption{The field of RX J0527 showing the stars (green) and the red and blue wings of H$\alpha$ . This shows that 
star 1a has extended subarcsecond H$\alpha$ emission.}
     \label{fig3}
  \end{figure}

Figure~\ref{fig3} shows the result of this procedure; star 1 is actually a blend of two quite similar stars named here as 1a (the brightest to the left) and star 1b, separated by $0.8\arcsec$. It is clear that spatially resolved H$\alpha$ emission is seen with subarcsecond structure that looks bipolar. The H$\alpha$ emission is extended but of about $0.3\arcsec$, which corresponds to 0.073 pc ($\sim3$ light months).

Star 1, defined as such by Greiner at al. (\cite{greiner96b}), was observed by Cowley et al. (\cite{cow97}), who found it to be a B8 IV with magnitude of $V= 17.3$; $B-V= 0.1$ and $U-B= -0.2$. They name it star 4. Our GMOS-IFU observations comprise stars 1, 9, 6, and 7 (Fig.~\ref{fig1}). The flux ratio in the $V$ band between 1a and 1b is 1.43 so, considering the apparent magnitude of both as determined by Cowley et al. (\cite{cow97}), star 1a has $V= 17.9$ and star 1b has $V= 18.3$, which yields absolute magnitudes of $-0.9$ and $-0.5$, respectively. We classify star 1a as B5e V. Such a star has a mass of $M_2 \sim 6$ M$_{\odot}$ and a radius of $R_2 \sim 3.8$ R$_{\odot}$. For this star to have a white dwarf companion, it would need a minimum orbital period of 21 h. We conclude that the 9.4 h period, found by Alcock et al. (\cite{alco}) and suggested by them as associated either to star 6 or 9, is probably not the orbital period of the supersoft X-ray source (SSS).

\begin{figure}
      \resizebox{\hsize}{!}{\includegraphics[clip]{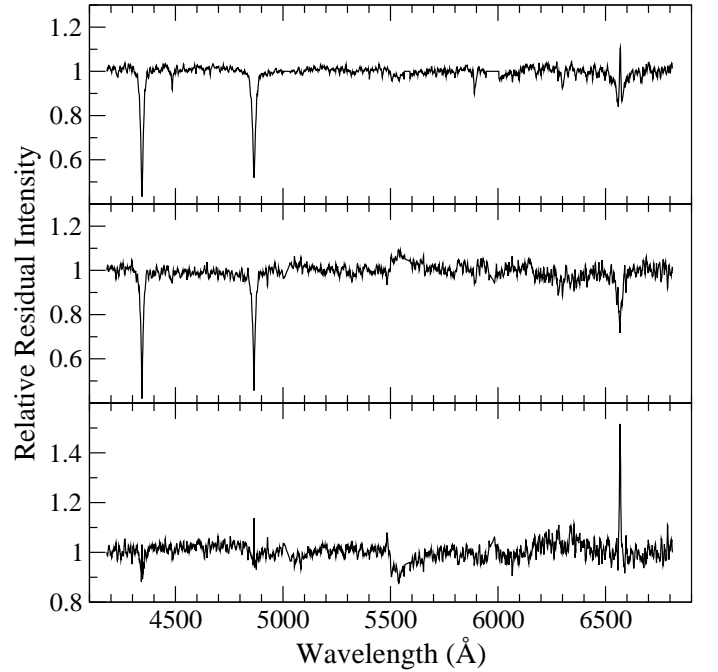}}
           \caption{Spectra of stars 1a (\textit{top}), 1b (\textit{center}), and the ratio of the spectra 1a to 1b (\textit{bottom}). The glitch at 5500 \AA{} is instrumental; two CCD gaps can be seen at 5000 \AA{} and 6000 \AA.}
     \label{fig4}
  \end{figure}

The spectra of stars 1a and 1b are shown in Fig.~\ref{fig4}. They appear very similar. The ratio of the two spectra enhances the narrow Balmer lines and H$\beta$ also appears, although no hint of it is visible in the original spectrum. Star 6 has $V\sim19.0$, and its optical spectrum is  about G III, with an absolute magnitude of $M_\mathrm{v} \sim +0.2$. Star 9 is a main sequence B star with $M_\mathrm{v} \sim-0.1$.

The narrow and weak H$\alpha$ that stands out from the broad absorption of star 1a has an equivalent width of
$W=-2.27~\AA{}$. This corresponds to a luminosity of $L_{\mathrm{H}\alpha} \sim1.6 \times 10^{32}$ 
erg~s$^{-1}$ ($\sim0.039$ L$_{\odot}$). Given the dimension of 0.073 pc, this implies (Osterbrock \cite{oster}) that its 
nebular density is 140~cm$^{-3}$  for a filling factor $f\sim1$. Photoionization models using CLOUDY show that 
this would produce too strong an emission of [\ion{O}{iii}] 4959 \AA{}. Notice that the [\ion{O}{iii}] 5007 \AA{} line 
is in the CCD gap. No forbidden line is, in fact, observed in the spectrum. A self-consistent photoionization model 
can be produced with an ionization temperature of T$\sim200\,000$ K, L$_\mathrm{bb}\sim 10^{34}$ erg~s$^{-1}$, a filling
 factor of $f\sim10^{-2}$, and an electron density of $N_\mathrm{e} = 10^3$~cm$^{-3}$. This model predicts weak
 [\ion{O}{iii}] emission, consistent with the observation.

\section{Discussion and conclusions} 

The main result of this work is the identification of the optical counterpart of RX J0527.8-6954 by the discovery of a bipolar H$\alpha$ subarcsecond extended emission. We found that star 1 is a blend of two stars with quite similar brightness and spectral types. Star 1a (the brightest one) is associated to the H$\alpha$ emission. The extended character of H$\alpha$ is  expected, as many CBSS show spectroscopic evidence of jets and winds and are supposed to be copious UV sources that certainly produce ionization cones.

If there is indeed a white dwarf companion, it must be quite massive, most likely above 1 M$_{\odot}$. Kahabka (\cite{kahab}) found a mass of $M_{\mathrm{wd}}$ = 1.14 to 1.34 M$_{\odot}$ and a temperature of 5 to 6$ \times10^5$ K, from both the decay and recurrence times of X-ray outbursts, but admitted a high uncertainty to these parameters. Fitting a model atmosphere to ROSAT X-ray spectra, Suleimanov \& Ibragimov (\cite{sule}) have derived a mass for the white dwarf of 1.2 to 1.4 M$_{\odot}$, one of the highest masses estimated among the CBSS. For star 1a to accommodate a white dwarf companion, their orbital period has to exceed 21 h. Given the uncertainties involved in the fit to the ROSAT data, one should also consider the possibility that the compact star is not a white dwarf. If it is a neutron star of higher mass, the orbital period could be shorter, perhaps as short as 9 h.

RX J0527.8-6954 is a transient SSS with a bolometric luminosity between $0.4\times10^{37}$ and $0.9\times10^{37}$ erg~s$^{-1}$ (Suleimanov \& Ibragimov \cite{sule}); the X-ray lightcurve showed a steady decay along 5 years (Greiner et al. \cite{greiner96b}). This differentiates the system from other known CBSS. A second difference is that the total optical brightness of the compact component, which usually dominates the optical emission, is not seen here. These  characteristics suggest that the system belongs to a distinct class of SSS. Three other SSS, in the LMC, SMC and M31 (Kahabka et al. \cite{kaha06}, Takei et al. \cite{takei}, Nelson et al. \cite{nelson}), may also be Be/WD binaries where the supersoft emission is due to stable nuclear burning of the white dwarf's envelope. This kind of sources may be associated to SSS observed in young stellar populations like the spiral arms of some galaxies. The supersoft X-ray emission of RX J0527.8-6954 was only detected a single time, no source being registered at its position in historical archives. This is also the case of the Be SSS transient detected in the SMC (Takei et al. \cite{takei}). In the present case the Be classification probably derives only from the bipolar emission and not from a circumstellar disc, as is usually the case in Be stars implying, perhaps, in a distinct nature for its accretion process.

\begin{acknowledgements}
Data obtained under Gemini program GS-2004B-Q-2. We would like to thank G. Ferland for the use of the CLOUDY program.
A.S. Oliveira (grant 03/12618-7), T.V. Ricci (grant 08/06988-0), R.B. Menezes (grant 08/11087-1) and B.W. Borges (grant 08/04530-6) acknowledge FAPESP -- Funda\c{c}\~{a}o de Amparo \`{a} Pesquisa do Estado de S\~{a}o Paulo -- for financial support. 

\end{acknowledgements}


\begin{thebibliography}{}

\bibitem[1997]{alco}
Alcock, C., Allsman, R.A., Alves, D.R. et al. 1997, MNRAS, 291, L13

\bibitem[2002]{alli}
Allington-Smith, J., Murray, G., Content, R. et al. 2002, PASP, 114, 892

\bibitem[1997]{cow97}
Cowley, A.P., Schmidtke, P.C., McGrath, T.K. et al. 1997, PASP, 109, 21 

\bibitem[1991]{greiner91}
Greiner, J., Hasinger, G., \& Kahabka, P. 1991, A\&A, 246, L17

\bibitem[1996]{greiner96}
Greiner, J. \& Hazen, L. M. 1996, IBVS, 4409

\bibitem[1996a]{greiner96a}
Greiner, J., Schwarz, R., Hasinger, G. \& Orio, M. 1996, A\&A, 312, 88

\bibitem[1996b]{greiner96b}
Greiner, J., Schwarz, R., Hasinger, G. \& Orio, M. 1996b, in Supersoft X-ray Sources, ed. J. Greiner, LNP, 472, 145

\bibitem[2004]{hook}
Hook, I.M., Jorgensen, I., Allington-Smith, J.R. et al. 2004, PASP, 116, 425

\bibitem[1995]{kahab}
Kahabka, P. 1995, A\&A, 304, 227

\bibitem[1997]{kah}
Kahabka, P., \& van den Heuvel, E.P.J. 1997, ARA\&A, 35, 69

\bibitem[2006]{kaha06}
Kahabka, P., Haberl, F., Payne, J.L. \& Filipovi\'c, M.D. 2006, A\&A, 458, 285

\bibitem[2010]{nelson}
Nelson, T., Orio, M. \& Di Mille, F. 2010, AN, 331, 223

\bibitem[1989]{oster}
Osterbrock, D. E. 1989, Astrophysics of Gaseous Nebulae and Active Galactic Nuclei, University Science Books, Mill Valey, CA

\bibitem[1995]{remi}
Remillard, R.A., Rappaport, S. \& Macri, L.M. 1995, ApJ, 439, 646

\bibitem[1998]{stei98}
Steiner, J.E. \& Diaz, M.P. 1998, PASP, 110, 276

\bibitem[2007]{stei07}
Steiner, J.E., Oliveira, A.S., Torres, C.A.O., \& Damineli, A. 2007, A\&A, 471, L25 

\bibitem[2003]{sule}
Suleimanov, V.F. \& Ibragimov, A. A. 2003, Astron. Rep., 47, 197

\bibitem[2008]{takei}
Takei, D., Tsujimoto, M., Kitamoto, S. et al. 2008, PASJ, 60, S231

\bibitem[1992]{heuvel}
van den Heuvel, E.P.J., Bhattacharya, D., Nomoto, K. \& Rappaport, S.A., 1992, A\&A, 262, 97

\bibitem[1998]{tees}
van Teeseling, A. \& King, A.R. 1998, A\&A, 338, 957

\bibitem[1991]{trumper}
Tr\"umper, J., Hasinger, G., Aschenbach, B., et al. 1991, \nat, 349, 579


\end{thebibliography}
\end{document}